\documentclass{aa}
\usepackage{graphicx}
\usepackage{natbib}
\usepackage{txfonts}
\bibpunct{(}{)}{;}{a}{}{,}    

\begin{document}
\title{Hanle effect in the CN violet system with LTE modeling}
\author{A. I. Shapiro \inst{1}  \and S. V. Berdyugina \inst{1,2}  \and D. M. Fluri \inst{1} \and J. O. Stenflo \inst{1,3}}
\offprints{A.I. Shapiro}

\institute{Institute of Astronomy, ETH Zurich, 8092 Zurich, Switzerland\\
\email{shapiro@astro.phys.ethz.ch}
\and Tuorla Observatory, University of Turku, 21500 Piikki$\ddot{\rm o}$, Finland
\and Faculty of Mathematics {\rm \&} Science, University of Zurich, 8057 Zurich, Switzerland}

\date{Received 11/06/2007; accepted 04/09/2007}

\abstract
{Weak entangled magnetic fields with mixed polarity occupy the main part of the quiet Sun. The Zeeman effect diagnostics fails to measure such fields because of cancellation in circular polarization. However, the Hanle effect diagnostics, accessible  through the second solar spectrum, provides us with a very sensitive tool for studying the distribution of weak magnetic fields on the Sun.}
{Molecular lines are very strong and even dominate in some regions of the second solar spectrum. The CN $B \, {}^{2} \Sigma - X \, {}^{2}
\Sigma$ system is one of the richest and most promising systems for molecular diagnostics and well suited for the application of the differential Hanle effect method. The aim is to interpret observations of the CN $B \, {}^{2} \Sigma - X \, {}^{2} \Sigma$ system using the Hanle effect and to obtain an estimation of the magnetic field strength.}
{We assume that the CN molecular layer is situated above the region where the continuum radiation is formed and employ the single-scattering approximation. Together with the Hanle effect theory this provides us with a model that can diagnose turbulent magnetic fields.}
{We have succeeded in fitting modeled CN lines in several regions of the second solar spectrum to observations and obtained a magnetic field strength in the range from 10--30 G in the upper solar photosphere depending on the considered lines.}
{}


\maketitle
%

\section{Introduction}\label{sec:intro}
The understanding of the solar surface magnetic field plays a key role in modern solar physics since the magnetic field is connected to and even controls most phenomena of solar activity. The solar magnetic field is structured in a fractal-like pattern with a significant degree of self-similarity over many scales \citep{stenfloholzreuter2003a}, while the field strength covers many orders of magnitude reaching up to a few kG. Observationally,  we can resolve only a small fraction of the magnetic structures, which can be accurately measured via the Zeeman effect diagnostics. However, about 99\% of the photospheric volume is filled by unresolved, entangled magnetic fields with mixed polarity, not accessible with the Zeeman effect due to cancellation of the polarimetric signal, that might contribute a substantial amount of magnetic energy to the solar photosphere \citep{trujilloetal2004}. 

These hidden magnetic fields can be accessed with the Hanle effect \citep{stenflo1982}, which modifies the properties of  coherent scattering processes. The Hanle effect can thus only be employed through the spectrum formed by the coherent scattering, which is called  the ``second solar spectrum'' \citep{ivanov1991} due to its richness and significance. The  
Hanle effect usually leads to  a reduction and rotation of line polarization as compared to the non-magnetic case.  The net-influence of rotation cancels out in a turbulent magnetic field, but the depolarization remains visible for most geometries even in a turbulent magnetic field. Moreover, the Hanle effect is extremely sensitive to the weak field. Therefore, the second solar spectrum provides us with a very
sensitive tool for studying the distribution of weak magnetic fields on the Sun and gives the possibility to measure spatially unresolved mixed polarity magnetic fields.

Molecular lines are particularly valuable for magnetic field diagnostics thanks to their broad range of magnetic sensitivities within narrow spectral regions. This allows us to employ the differential Hanle effect \citep{stenfloetal1998}, which greatly reduces the model dependence of deduced magnetic field strengths, and has lead to the first unambiguous detection of molecular Hanle effect in the second solar spectrum \citep{fluriberdyugina2004b}. Recent studies of the molecular Hanle effect have yielded field strengths of the order of 10 G for the quiet solar photosphere, assuming a single-valued, isotropic, turbulent magnetic field \citep{faurobertarnaud2003, berdyuginafluri2004, trujilloetal2004}, although larger field strengths are possible depending on the assumed collision rate, in particular when interpreting MgH lines \citep{asensiotrujillo2005, bommieretal2006}.

Diagnostics with molecular Hanle effect has concentrated so far only on two molecules, $\rm{C}_2$ and MgH. It is very important to extend the technique to other molecules because in general different molecules form at different heights due to their large temperature sensitivity and can thus sample other layers of the solar atmosphere. Combined with atomic Hanle effect, which is well established since the work by \citet{faurobert1992}, only such a multitude of lines will allow us to constrain the 3-dimensional structure of the unresolved solar magnetic field. 

In the present paper we apply the Hanle effect to a new molecule, namely the CN violet system, for which we have recently developed a complete theory of the Hanle effect taking into account the Paschen-Back effect and interference between doublet components \citep{shapiroetal2006b}. After describing our radiative transfer method we fit our model to observations in three spectral regions of the second solar spectrum, which enables us to determine the magnetic field strength in the quiet solar photosphere.

\section{Radiative transfer model}\label{sec:RT}
In this section we introduce our method for solving the radiative transfer problem. It is based on the modeling strategy developed by \citet{faurobertarnaud2003}, who have calculated scattering polarization at the line centers of ${\rm C}_2$ and MgH transitions. Here, we expand their method to obtain wavelength dependent spectra not restricted to line centers.

The radiative transfer model is optimized for relatively weak lines whose scattering polarization is calculated within the framework of a plane-parallel atmosphere using the single-scattering approximation. The CN lines are assumed to form in a homogeneous atmospheric layer situated above the continuum formation region. We fully account for line blending and for depolarization of the continuum polarization by line absorption and scattering. In this respect, the current model is more complex than the one employed by \citet{berdyuginafluri2004} for very weak ${\rm C}_2$ lines. The CN bands that are modeled in this paper include stronger lines and require a consistent treatment of continuum depolarization, although we increase at the same time the model dependence of the magnetic field diagnostics, because we rely on a proper computation of the continuum polarization. Nonetheless, the model remains simple and ideally suited for differential Hanle effect diagnostics.

First we introduce the necessary notation and the source function. Then we discuss in detail the two-layer approximation, the free model parameters, and the fitting procedure for comparing the model with observations.

\subsection{Source function}\label{subsec:SF}
The total, monochromatic optical depth is given by
\begin{equation}
d  \tau (\nu) = - \left( \alpha_{\rm c} + \sum_{i=1}^{n} \alpha_{\ell}^{i} \varphi^{i} (\nu)  \right ) \, dz,
\label{eq:depth}
\end{equation}
where $\alpha_{\rm c}=k_{\rm c}+\sigma_{\rm c}$ is the continuum opacity (with $k_{\rm c}$ and $\sigma_{\rm c}$  the continuum absorption and scattering coefficients), $\alpha_{\ell}^{i}=k_{\ell}^{i} + \sigma_{\ell}^{i}$ is the integrated line opacity in the {\it i}-th line (with $k_{\ell}^{i}$ and ${\sigma}_{\ell}^{i}$ the line absorption and scattering coefficients), and $\varphi^{i} (\nu)$ is the normalized  Voigt profile function. The opacities in molecular lines depend on the main line parameters and on the temperature of the CN layer $T_{\rm eff}$
\begin{equation}
\alpha_{\ell}^{i} \propto g_{\ell}^{i} \, f_{\ell}^{i} \exp(-E_{\ell}^{i}/kT_{\rm eff}), 
\label{eq:opacity}
\end{equation}
where $g_{\ell}^{i}$ and $E_{\ell}^{i}$ are the statistical weight and energy, respectively, of the lower level, and  $f_{\ell}^{i}$ is  the absorption oscillator strength, which can be calculated with the pure Hund's case~(b) expressions  \citep{schadee1978, whitingnicholls1974} because the
orbital angular momenta for both the upper and lower states are equal to zero in the CN violet system.

Now we can write the radiative transfer equation in the usual form
\begin{equation}
\mu \frac{\partial \vec{I}}{\partial \, \tau} = \vec{I}-\vec{S},   
\label{eq:tq}
\end{equation}
where $\vec{S} \equiv \vec{S} (\tau, \nu, \mu)$ is the total source function 
\begin{equation}
{\vec{S}}  = \frac{k_{\rm c} {\vec{B}}_{\rm th} +\sum\limits_{i=1}^{n} k_{\ell}^i \varphi^{i} (\nu) {\vec{B}}_{\rm th} + \sigma_{\rm c} {\vec{S}}_{\rm c} + \sum\limits_{i=1}^{n} \sigma_{\ell}^i \varphi^{i} (\nu) {\vec{S}}_{\ell}^i}{\alpha_{\rm c}+ \sum\limits_{i=1}^{n} \alpha_{\ell}^i \varphi^{i} (\nu) },
\label{eq:totalS}
\end{equation}
with $\vec{B}_{\rm th}$ the Planck function,  $\vec{S}_{\rm c}   \!\! \equiv  \!\! \vec{S}_{\rm c} (\tau, \nu, \mu)$ the continuum source function, and  $\vec{S}_{\ell}^{i} \! \equiv \! \vec{S}_{\ell}^{i} (\tau, \nu, \mu)$ the scattering part of the line source function of the {\it i}-th line.

In our further discussion we employ the single-scattering approximation, since the CN lines for which we will apply this theory are optically thin. Then, the scattering part of the line source function takes the form
\begin{equation}
{\vec{S}}_{\ell}^{i}  = \frac{1}{2} \int\limits_0^1  {\vec P}_{\ell}^{i} (\mu, \mu', W_2^{i}, W_H^{i}) {\vec I}_{\rm{inc}} (\mu') 
\exp(-\frac{{{\tau}_{\rm{inc}}}(\nu)-\tau}{\mu'}) \, d \mu',
\label{eq:lineS}
\end{equation}
where $\vec{I}_{\rm inc} \!\! = \!\!  \vec{I} (\tau_{\rm inc}, \nu, \mu)$ is the Stokes vector at the lower boundary with an optical depth ${\tau}_{\rm inc}$. 

Within the weak-field approximation and under the assumption of an isotropic turbulent magnetic field  the phase matrix $\vec{P}_{\ell}^{i}$ is parametrized as
\begin{equation}
{\vec P_{\ell}^{i}} = {\vec E}_{11}+\frac{3}{4} \, W_2^{i} W_{\rm H}^{i}  \, {\vec P}^{(2)}.
\label{eq:P}
\end{equation}
The matrix $\vec{E}_{11}$ represents isotropic, unpolarized scattering, while ${\vec P}^{(2)}$ describes linearly polarized coherent scattering \citep[e.g.][]{stenflo1994}. The line polarizability $W_2^{i}$, calculated as in \citet{berdyuginaetal2002b}, corresponds to the branching ratio of polarized and unpolarized scattering. The physics of the Hanle effect is contained in the Hanle depolarization factor

\begin{equation}
W_{\rm H}=1-0.4  \left (        \frac {   \gamma_{\rm H}^2   } {1+\gamma_{\rm H}^2} +   \frac { 4  \gamma_{\rm H}^2   } {1+4 \gamma_{\rm H}^2}            \right ),
\label{eq:WH}
\end{equation}
with
\begin{equation}
\gamma_{\rm H}=0.88 \, \frac{g_{\rm L} B}{\Gamma_{\rm R} + D^{(2)} }.
\label{eq:gamma}
\end{equation}
The variables $\Gamma_{\rm R}$ and $D^{(2)}$ are the radiative damping and depolarizing collision rates, respectively, and $B$ is the strength of the  single-valued, isotropic, microturbulent magnetic field. For the analysis in this paper we neglect depolarizing collisions. In the CN violet system we are dealing with ${}^2 \Sigma-  {}^2 \Sigma$ transitions, so that the effective  Land\'{e} factors $g_{\rm L}$ of the upper state of the two doublet components can be expressed through the quantum number of the total angular momentum $J$ or, alternatively, of the angular momentum $N$, i.e. the total angular momentum without spin,

\begin{equation}
\left. 
\begin{array}{l}
g_{\rm L1}={1}/({N+\frac{1}{2}})={1}/{J},     \\
g_{\rm L2}=-{1}/({N+\frac{1}{2}})=-{1}/({J+1}), \\
\end{array}
\right.
\label{eq:Lande}
\end{equation}
where the indices $1$ and $2$ refer to the first and second doublet components with $J=N-1/2$ and with $J=N+1/2$, respectively.

With our choice of the phase matrix we implicitly neglect interferences between different fine structure components and perturbations by the Paschen-Back effect. The latter becomes relevant in the CN violet system only for field strengths exceeding 100 G \citep{shapiroetal2006b}, while the fields detected in the present study are weaker than 30 G.

Let us rewrite Eq.~(\ref{eq:totalS}) in the form
\begin{eqnarray}
{\vec{S}}= && \!\!\! \frac{\sigma_{\rm c}+ \sum\limits_{i=1}^{n} \sigma_{\ell}^i \varphi^{i} (\nu)}{\alpha_{\rm c}+ \sum\limits_{i=1}^{n} \alpha_{\ell}^i \varphi^{i} (\nu)}
\left ( \sum\limits_{i=1}^{n} \chi_i {\vec{S}}_{\ell}^i + \chi_{\rm c} {\vec{S}}_{\rm c} \right )  \nonumber \\
&& \!\!\!\!\! + \,\, \frac{k_{\rm c}+ \sum\limits_{i=1}^{n} k_{\ell}^i \varphi^{i} (\nu)}{\alpha_{\rm c}+ \sum\limits_{i=1}^{n} \alpha_{\ell}^i \varphi^{i} (\nu)} {\vec{B}}_{\rm th}, 
\label{eq:lineSr}
\end{eqnarray}
where the coefficients  $\chi_i$ and $\chi_{\rm c}$ correspond to the conditional probabilities that scattering occurs in the {\it i}-th line or in the continuum, respectively. In general they depend on frequency and are defined as
\begin{equation}
\chi_i=\frac{\sigma_{\ell}^i \varphi^{i} (\nu) }{\sigma_{\rm c}+\sum\limits_{j=1}^{n} \sigma_{\ell}^j \varphi^{j} (\nu)}, ~~~~
\chi_{\rm c}=\frac{\sigma_{\rm c} }{\sigma_{\rm c}+\sum\limits_{j=1}^{n} \sigma_{\ell}^j \varphi^{j} (\nu)}.
\label{eq:conprob}
\end{equation}
The emergent Stokes vector is given by the formal solution of the radiative transfer equation
\begin{equation}
{\vec{I}}_{\rm em} (\nu, \mu)=\int\limits_0^{ {\tau}_{\rm inc} (\nu) } {\vec S} (\tau, \nu, \mu) {\rm e}^{-{\tau}/{\mu}} \, \frac{d \tau}{\mu} + \vec{I}_{\rm inc} (\nu, \mu) 
{\rm e}^{-{{\tau}_{\rm inc} (\nu) }/{\mu}}. 
\label{eq:Ingeneral}
\end{equation}
\subsection{Two-layer approximation}\label{subsec:twolayer}
We assume that the CN molecular layer is situated above the region where the continuum radiation is formed. The CN layer is isothermal with a homogeneous CN density, and its lower boundary is irradiated by the solar continuum radiation. While passing through the molecular layer this initial incident radiation can be scattered (but only once, as we adopt the single scattering approximation) or absorbed by CN molecules.

The ``continuum'' layer corresponds to the lower photosphere of the average quiet Sun, defines the lower boundary condition of the CN layer, and is thus crucial for the formation of the CN lines. In particular, the continuum layer determines the anisotropy of the continuum radiation, without which scattering polarization would cancel out, and it sets the degree of continuum polarization, which is subsequently depolarized by the line opacity.

In our modeling strategy we take the continuum intensity and polarization from previous work, rather than recalculating these values. The continuum intensity is obtained from the analytical expressions for limb-darkening derived from observations of the quiet Sun \citep{neckel1996}. We assume the radiation field to be azimuthally symmetric. The values of the continuum polarization correspond to the calculations by \citet{stenflo2005} for $\mu\!=\!0.1$. The continuum polarization for arbitrary $\mu$-values are computed by scaling the value at $\mu\!=\!0.1$ with the analytical expressions introduced by \citet{fluristenflo1999a}, which describe the $\mu$-dependence of the continuum polarization and are based on theoretical modeling of the quiet solar photosphere.

The CN layer is described by the source function and the radiative transfer equation given in Sect.~\ref{subsec:SF}. In Eqs.~(\ref{eq:lineS}) and (\ref{eq:Ingeneral}) we insert the Stokes vector $\vec{I}_\mathrm{inc}$, which is incident into the CN layer at the lower boundary and given by the frequency and angle dependent continuum intensity and polarization formed in the continuum layer. The same equations require also the total optical thickness $\tau_\mathrm{inc}\!=\!{\cal T}_\mathrm{CN}$ of the CN layer. Assuming $\alpha_\mathrm{c}\!=\!0$, since the continuum is formed below the CN layer, the total optical thickness of the CN layer ${\cal T}_\mathrm{CN}$ can be calculated with the expression
\begin{equation}\label{eq:totaltau}
{\cal T}_\mathrm{CN}\,(\nu)
= {\cal T}_\mathrm{norm}  \cdot \sum_{i=1}^{n}\alpha_{\ell}^{i}\phi^{i}(\nu)
\rm \ ,
\end{equation}
where the frequency independent coefficient ${\cal T}_\mathrm{norm}$ is a free parameter of the model. It depends on the CN density and the geometrical thickness of the CN layer.

Within the two-layer approximation the total source function, Eq.~(\ref{eq:lineSr}), is further simplified since the opacity in the CN layer is only due to CN molecules, i.e.\ $k_\mathrm{c}\!=\!\sigma_\mathrm{c}\!=\!0$.  We assume that the ratio of the scattering and absorption probabilities $c_\mathrm{sc}\!=\!\sigma_{\ell}^{i}/\alpha_{\ell}^{i}$ is the same for all molecular lines. This is basically equivalent to the assumption that the inelastic collision rate is independent of the involved transitions because the life time of the exited CN state $B^2\Sigma$ is almost independent of its quantum numbers (with the exception of some states with very small $J$ numbers). Furthermore we neglect in our model the elastic collisions. Equation~(\ref{eq:lineSr}) is then reduced to the form
\begin{equation}\label{eq:CNsource}
\vec{S} = 
c_\mathrm{sc}\left( \sum_{i=1}^{n} \chi_{i} \vec{S}_{\ell}^{i} \right)
+ \left( 1-c_\mathrm{sc} \right) \vec{B}_\mathrm{th}
\rm \ .
\end{equation}
Here the first term with the summation represents the scattered diffuse part of the radiation field, restricted within the CN layer to line contributions, while the second term corresponds to the thermal part.

\begin{figure*}
\centering
\includegraphics[height=9.5cm]{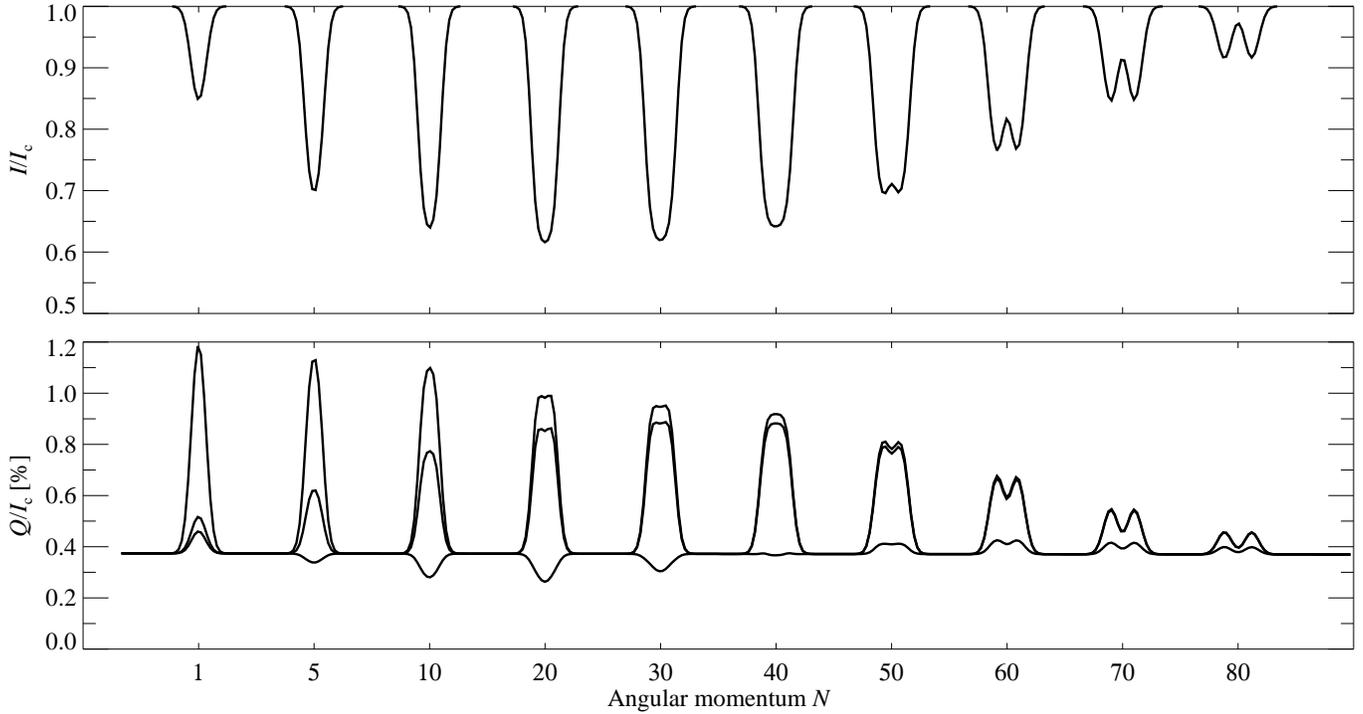}
\caption{Differential Hanle effect in synthetic Stokes profiles of R-branch lines in the CN violet (0,0) band. For selected angular momenta $N$ of the lower state we show the complete doublet, which causes the double peak structure for large $N$.
Three values for the turbulent magnetic field strength were chosen, giving identical Stokes $I$ profiles but large modifications in the Stokes $Q$ profiles: 0\,G (upper curve), 10\,G (intermediate curve), and 100\,G (lower curve). In this and all other figures we assume an elastic collision rate  equal to zero.}
\label{fig:spectrum_ex}
\end{figure*}
\begin{figure}
\resizebox{\hsize}{!}{\includegraphics{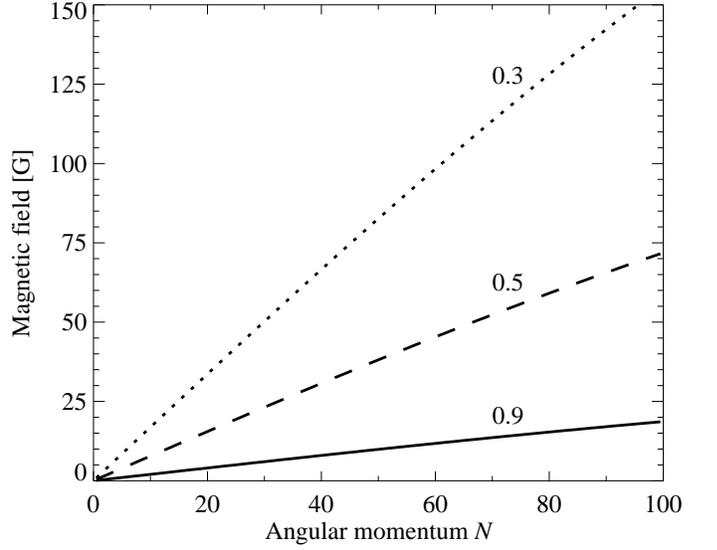}}
\caption{The dependence of the Hanle depolarization factor $W_{\rm H}$, given in Eq.~(\ref{eq:WH}), on the magnetic field strength and the $N$ quantum number for the CN violet (0,0) band. The three curves correspond to the depolarization factors $W_{\rm H}$ equal to 0.9 (solid), 0.5 (dashed), and 0.3 (dotted), the last of which is already close to the asymptotic saturation limit of 0.2.}
\label{fig:depol}
\end{figure}

The emergent Stokes vector is calculated employing the formal solution of the radiative transfer equation. To improve the accuracy of our results we divide the CN layer into $k$ sublayers with equal optical thickness. The source function is assumed to be constant within each sublayer and evaluated at the center of each layer, giving us $k$ different values $\vec{S}_1$\ldots$\vec{S}_k$. Introducing the above assumptions into Eq.~(\ref{eq:Ingeneral}), we find for the emergent radiation field at a given frequency
\begin{equation}\label{eq:sublayers}
\vec{I}(\mu) = 
\vec{I}_\mathrm{inc}(\mu) \mathrm{e}^{-{\cal T}_\mathrm{CN}/\mu}
+\sum_{j=1}^{k} \vec{S}_{j}(\mu) 
                \left( \mathrm{e}^{-{\cal T}_\mathrm{CN}(j-1)/k\mu}
                       - \mathrm{e}^{-{\cal T}_\mathrm{CN}j/k\mu}
                \right)
\rm \ .
\end{equation}
The first term describes the radiation transmitted through the CN layer while the summation in the second term adds up the contributions from the different CN sublayers to the emergent spectrum. Note that the depth dependence of the source function within the CN layer is introduced by the exponential factor in Eq.~(\ref{eq:lineS}), which describes absorption of the incident radiation and thus the height variation of the radiation field even within the isothermal slab. The scattering coefficient, on the other hand, depends only on the probability of the collisional de-excitation of the upper CN states, which under our assumptions remains constant for the whole molecular layer.

The subdivision of the molecular layer, expressed by Eq.~(\ref{eq:sublayers}), is a further sophistication of the model introduced by \citet{faurobertarnaud2003}, who have assumed one single value of the source function within the molecular layer. The sublayers take care of the depth dependence of the source function within the CN layer, which results from the optical depth $\tau$ in the exponent of Eq.~(\ref{eq:lineS}). Note that this does not contradict to our assumption of an isothermal CN layer with constant density. Our test calculations have shown that the emergent Stokes vector becomes independent of the number of sublayers for $k$ greater than five.

\subsection{Free parameters}\label{subsec:parameters}

The model described above contains several free parameters, which have to be determined by fitting the theoretical spectrum to observations. These are the temperature of the isothermal CN layer $T_\mathrm{eff}$, the scattering probability $c_\mathrm{sc}$, the optical depth normalization coefficient ${\cal T}_\mathrm{norm}$, and the magnetic field strength $B$.

The temperature $T_\mathrm{eff}$ affects the emergent spectrum in two ways. First of all, it enters via thermal term $\vec{B}_\mathrm{th}$ of the source function (cf.\ Eq.~(\ref{eq:CNsource})). This strongly influences the intensity, but not the other Stokes parameters, because of the strong temperature dependence of the Planck function and because the contribution from the thermal part is relatively large in the case of CN lines. Secondly, the line opacities depend on temperature via the Boltzmann factor (cf.\ Eq.~\ref{eq:opacity}), which affects the whole Stokes vector, although in a much weaker way than the thermal term.

The scattering probability $c_\mathrm{sc}$ defines the balance between the thermal and nonthermal parts of the total source function and influences the whole Stokes vector. The contribution to the Stokes vector from the CN layer linearly depends on $c_\mathrm{sc}$.

The optical depth normalization coefficient ${\cal T}_\mathrm{norm}$ describes the relative contributions to the Stokes vector from the CN layer and the continuum layer. It effectively determines the optical thickness of the CN layer, thus scaling both the molecular line depths in the intensity spectrum and the degree of line polarization.

Our main goal is the determination of the magnetic field strength $B$ assuming a turbulent, isotropic field distribution within the CN layer. The magnetic field enters via Eq.~(\ref{eq:gamma}) into the line source function. Since we consider weak fields and neglect Zeeman broadening, the magnetic field strength has virtually no effect on Stokes $I$ and only causes differential Hanle effect in the second solar spectrum (see Sec.~\ref{subsec:hanle}).

Usually, the influence of $B$ on the $Q/I_\mathrm{c}$ spectrum can be easily distinguished from the effects of $c_\mathrm{sc}$ and ${\cal T}_\mathrm{norm}$. The turbulent magnetic field causes a depolarization of the CN lines. The same can in principle also be obtained with a smaller scattering probability $c_\mathrm{sc}$. Even an increased ${\cal T}_\mathrm{norm}$ parameter can mimic the same behavior because it leads to stronger continuum depolarization. The crucial point, however, is that $c_\mathrm{sc}$ and ${\cal T}_\mathrm{norm}$ affect all molecular lines with comparable strength, while the magnetic field leads to differential depolarization, because the Hanle effect efficiency strongly depends on the $J$ quantum number (Fig.~\ref{fig:spectrum_ex}). Thus, the relative strength of CN lines in the second solar spectrum mainly depends on the magnetic field, but not so much on the other free parameters. 
Therefore, it is possible to diagnose the magnetic field strength when fitting several lines with different Hanle effect sensitivities.


\subsection{Fitting procedure}\label{subsec:fitting}
We simultaneously fit the modeled polarization spectrum to observed $Q/I_\mathrm{c}$ data and the modeled intensity spectrum to $I/I_\mathrm{c}$ observations. During the fitting procedure we vary the magnetic field strength $B$, the scattering probability $c_\mathrm{sc}$, the effective temperature $T_\mathrm{eff}$ and the optical depth normalization coefficient ${\cal T}_\mathrm{norm}$. The most suitable values for these free parameters are determined with a least-square method. More precisely, we minimize the quantity

\begin{eqnarray}\label{eq:chi}
\chi^2 = && \!\!\!
  \frac{\zeta}{n}
\sum_{i=1}^{n}{ \left( \frac{ (Q/I_\mathrm{c})_\mathrm{obs} 
                              - (Q/I_\mathrm{c})_\mathrm{th} }
                            { \sigma_{Q/I_\mathrm{c}} }  \right)^{2} }  \nonumber \\
                      \!\!\!  && \!\!\!\! +   \,\, \frac{(1-\zeta)}{n}
 \sum_{i=1}^{n} {  \left(   \frac{ (I/I_\mathrm{c})_\mathrm{obs} 
                              - (I/I_\mathrm{c})_\mathrm{th} }
                            { \sigma_{I/I_\mathrm{c}}  } \right)^{2} }  
\rm \ ,
\end{eqnarray}
where $n$ is the number of data points and $\sigma_{Q/I_\mathrm{c}}$ and $\sigma_{I/I_\mathrm{c}}$ are the absolute errors of the observed polarization and intensity, which we set to  0.008\% and 0.005\%, respectively. The coefficient  $\zeta$, with possible values in the range from $0$ to $1$, allows us to give larger weight either to the intensity or polarization fit. 
As the signal in polarization is smaller and  as only polarization is affected by the magnetic field
we overweight the fit to the polarization fit by choosing $\zeta=7/8$ (but the result of the fitting procedure is almost independent on the assumed $\zeta$ value).

The $\chi^2$ value in Eq.~(\ref{eq:chi}) is a measure of the discrepancy between observations and the theoretical model, and describes the accuracy of the fit. Wavelength points that are heavily affected by atomic blends  are excluded from the summation and the fitting procedure since we neglect atomic line opacity in our model.


\section{Diagnostics with the CN violet system}\label{sec:CN}

In this section we apply the previously developed theory to the interpretation of observations of the CN $B^2\Sigma-X^2\Sigma$ system, the so-called violet system, in several spectral regions, with the goal to determine the turbulent magnetic field strength in the quiet Sun. From the discussion in Sect.~\ref{subsec:parameters} it is clear that suitable wavelength regions should contain molecular lines with different Hanle effect sensitivities. Therefore, we first investigate the differential Hanle effect in the CN violet system. This will guide us to choose three ideal spectral windows in different molecular bands, which are then analyzed and discussed.


\subsection{Differential Hanle effect}\label{subsec:hanle}

Figure~\ref{fig:spectrum_ex} illustrates the different Hanle effect sensitivities of synthetic line profiles for the example of the CN violet (0,0) band (other CN violet bands give the same qualitative results). The relative degree of line polarization in the non-magnetic case is defined by three factors: the oscillator strength (which increases with the $J$ number), the scattering polarizability $W_2$ (which decreases with $J$ and asymptotically approaches a value of 0.1), and the Boltzmann factor (which also decreases with $J$). However, the final polarization obtained within the lines not only depends on the intrinsic line polarization. Rather it results from a combination of different contributions, namely the continuum polarization, continuum depolarization due to the line opacity, and line polarization, where only the latter is subject to the Hanle effect. It is clearly visible in Fig.~\ref{fig:spectrum_ex} that stronger lines lead to additional continuum depolarization \citep[cf.][]{fluristenflo2003}. Furthermore, we can identify the expected diverse behavior of the lines in a magnetic field. While the intensity profiles remain unaffected, the line polarization is depolarized with increasing field strength, reflecting the $J$ dependence of $W_\mathrm{H}$. A magnetic field of 10~G already leads to a nearly saturated Hanle effect in the lines with $N\!=\!1$ and~5, while the polarization in the $N\!=\!70$ and~80 lines remains almost unchanged compared to the non-magnetic case. The latter lines require a magnetic field strength in the order of 100~G for a considerable depolarization.

\begin{figure*}
\includegraphics{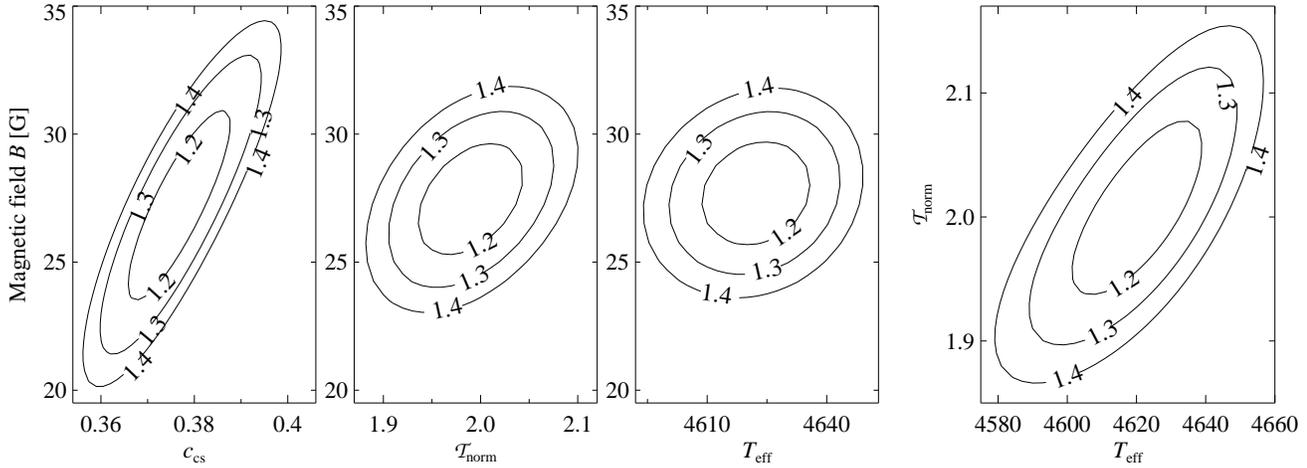}
\caption{${\chi}^2$-- contours of the fitting procedure in the first wavelength region ($ 3865.9\!\!-\!\!3866.8 $ {\AA}). The panels show 2-dimensional cuts in the parameter space through the minimum ${\chi}_{\rm min}^2=1.13$.  Numbers on the contours indicate the values of ${\chi}^2$ as
defined in Eq.~(\ref{eq:chi}).}
\label{fig:acc1}
\end{figure*}
\begin{figure}
\includegraphics[height=9.2cm]{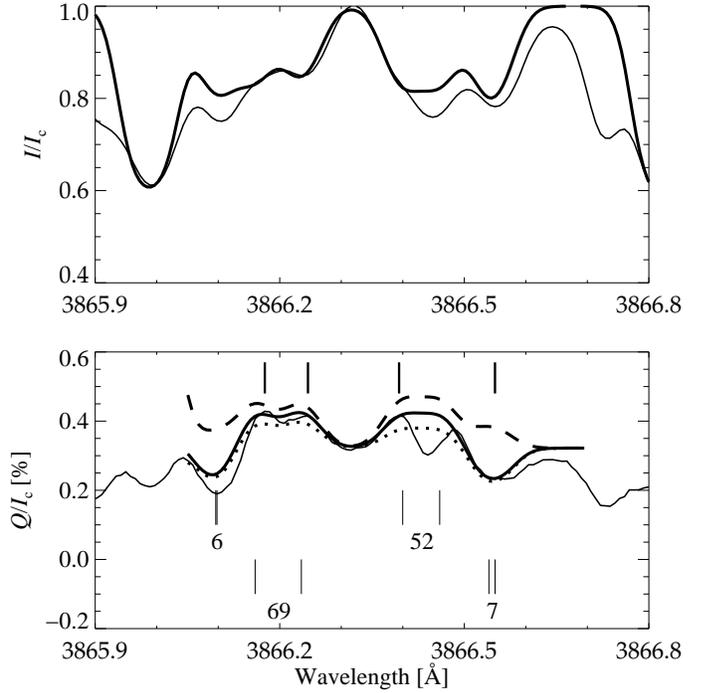}
\caption{Fits to the observations (thin solid  lines) of Stokes $I/I_{\rm c}$ and $Q/I_{\rm c}$ in the first wavelength region ($ 3865.9\!\!-\!\!3866.8 $ {\AA}). The assumed magnetic field strengths in the calculated spectra are 0 G (no depolarization, thick dashed line), 27 G (best fit, thick solid line), and 45 G (too much depolarization, dotted line).
The labels at the bottom of the lower panel indicate the position of the lines in the CN doublets with the corresponding $N$ quantum number of the lower states. The labels at the top of the lower panel indicate the wavelength points which were chosen for the fitting procedure.}
\label{fig:spectrum_ex1}
\end{figure}

The magnetic field necessary for sizable depolarization by the Hanle effect is shown in Fig.~\ref{fig:depol} for different $N$ quantum numbers, again for the CN violet (0,0) band. The different curves correspond to three values of $W_\mathrm{H}$ and identify the magnetic field range in which a specific line is sensitive to the Hanle effect. For example, lines with $N\!=\!40$ react to the Hanle effect mainly in the field range from about 10~G to 60~G. For stronger fields these lines are maximally depolarized. On the other hand, their Hanle effect depolarization is negligible for weaker fields, which makes them ideal reference lines below 10~G when compared to lines with much smaller $N$.

The ideal choice of spectral windows thus depends on the actual magnetic field strength. Based on previous results we selected narrow wavelength regions containing lines with $N$ greater than 40 and others with $N$ around 10. An additional selection criterion is a small number of atomic or molecular blends since they would significantly alter the intensity and polarization spectra.

\begin{figure*}
\includegraphics{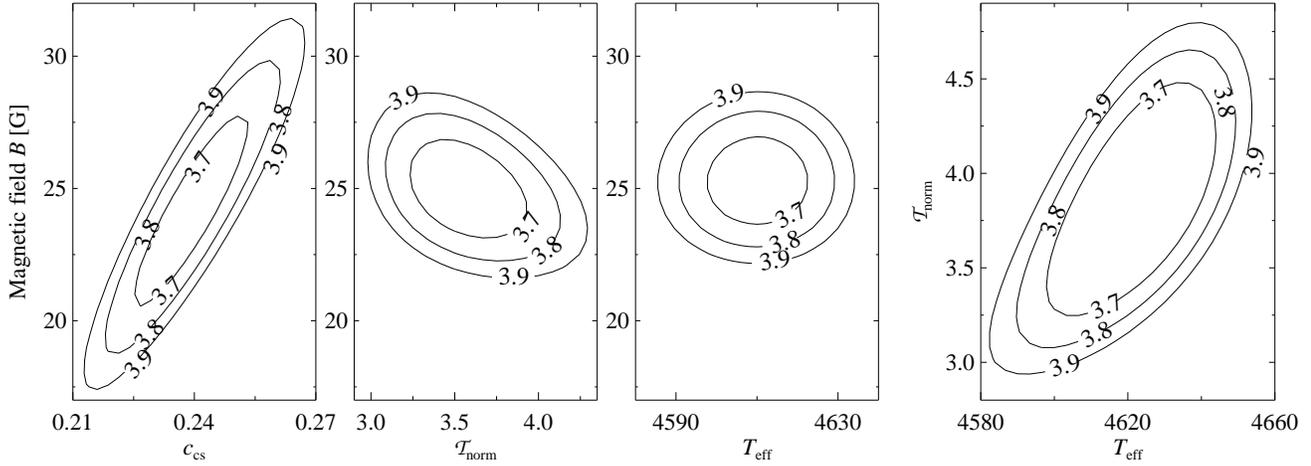}
\caption{${\chi}^2$-- contours as in Fig.~\ref{fig:spectrum_ex1} but for the second wavelength region close to the (1,1) bandhead. We obtain a minimum value  ${\chi}_{\rm min}^2=3.62$.}
\label{fig:acc2}
\end{figure*}

\begin{figure*}
\centering
\includegraphics[height=9.7cm]{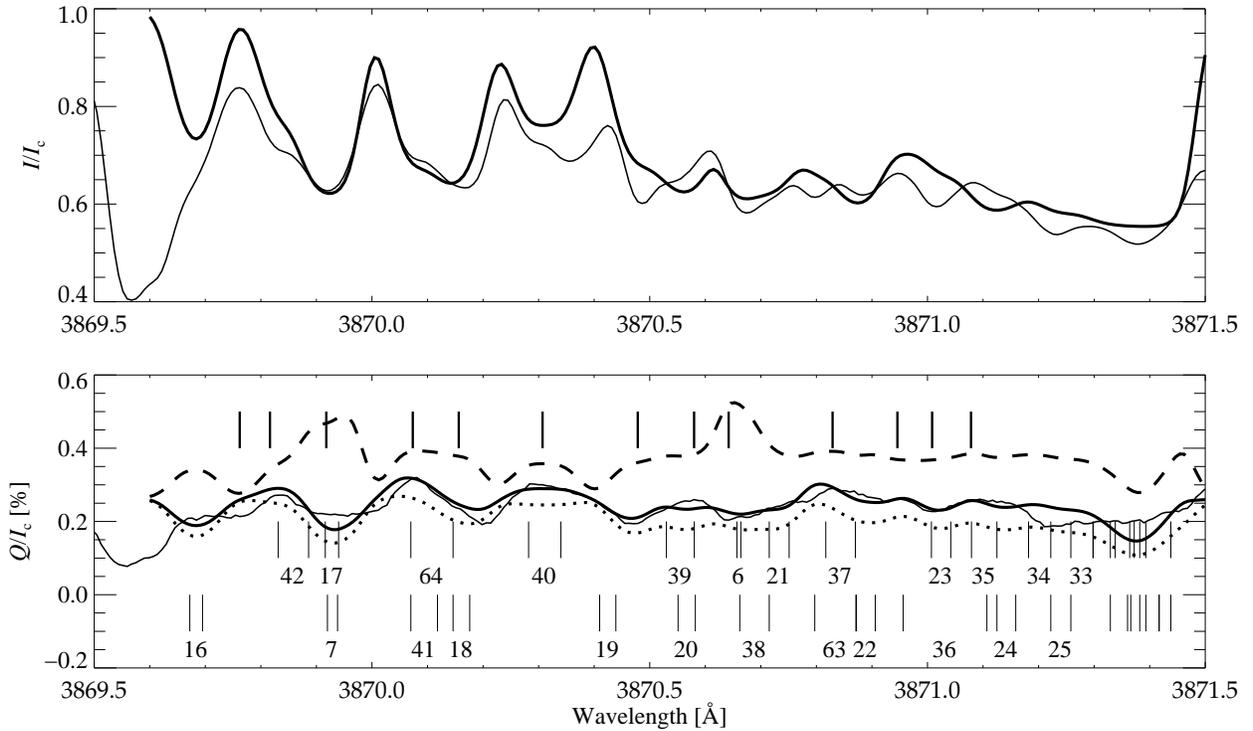}
\caption{Fits to the observations (thin solid  lines) of Stokes $I/I_{\rm c}$ and $Q/I_c$ in the second wavelength region close to the (1,1) bandhead. The assumed magnetic field strengths in the theoretical spectra are 0 G (no depolarization,  dashed line), 25 G (best fit, thick solid line) and 45 G (too much depolarization, dotted line). Labels at the bottom of the lower panel indicate the $N$ quantum numbers of the lower states and the line positions within each doublet (close to the bandhead $N$ is omitted due to overlap). The labels at the top of the lower panel indicate the wavelength points which were selected for the fitting procedure.}
\label{fig:spectrum_ex2}
\end{figure*}

\subsection{Observational data}\label{subsec:data}
We compare our model calculations with the data from the second solar spectrum atlas compiled by \citet{gandorfer2005atlas}. All spectra were recorded with the UV version of the Zurich Imaging Polarimeter ZIMPOL II \citep{gandorferetal2004} at the 1.5 m McMath-Pierce facility of the National Solar Observatory on Kitt Peak (Arizona, USA). The spectrograph slit was set parallel to the solar limb at the distance corresponding to the cosine of the heliocentric angle $\mu\!=\!0.1\pm0.02$. Due to the telescope guiding and seeing, the value of $\mu$ was not completely stable, resulting in a small uncertainty of the exact limb distance. Because the scattering polarization is very sensitive to small variations in $\mu$ very close to the solar limb, $\mu$ behaves like a semi-free parameter for our analysis within strict limits defined by the error bars.

One of the main problems for our fitting procedure is the uncertainty of the zero point of the polarization scale in the observations. To solve this problem, we selected several wavelengths as continuum reference points, which are far away from very strong lines and least affected by blends. Based on the theoretical continuum polarization \citep{stenflo2005}, we shifted the polarization scale of the observations appropriately. The same wavelength points were used to renormalize the $I/I_\mathrm{c}$ data.

We have chosen three wavelength regions for our calculations. These regions were not observed simultaneously and correspond, in principle, to different areas of the solar surface. Therefore measured parameters can vary from region to region.


\subsection{Region 1: diagnostics near 3866~\AA}\label{subsec:region1}

The first wavelength region  3865.9--3866.8~\AA\ contains four CN violet system doublets with $N\!=\!6$,~7 and 52 from the (1,1) vibrational band and with $N\!=\!69$ from the (0,0) vibrational band. 
Because of the relatively large Land\'e factors $(g_N\sim1/N)$, the lines with $N\!=\!6$ and 7 are very sensitive to the magnetic field and reach the saturated Hanle effect already at 5--10 G, while the lines with $N\!=\!52$ and 69 are much less affected by the magnetic field (see Fig.~\ref{fig:depol}). Therefore, this region is ideal for applying the differential Hanle effect technique, in particular since it contains lines with small and large $N$ quantum numbers from the same vibrational band, as the relative strength of the lines from different bands can be influenced by other factors. Furthermore, all lines can be measured simultaneously in a single observation because they are spread over less than 1 \AA.
\begin{figure*}
\centering
\includegraphics{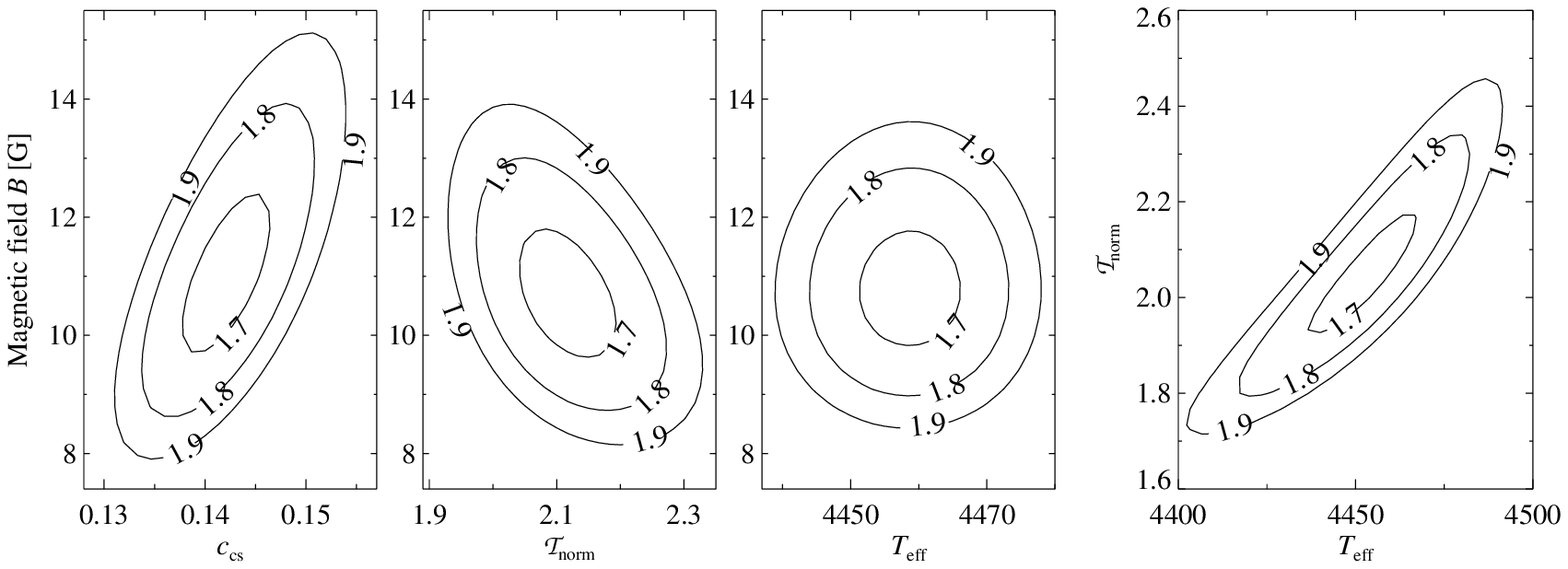}
\caption{${\chi}^2$-- contours as in Fig.~\ref{fig:spectrum_ex1} but for the third wavelength region near the (0,0) bandhead. We obtain a minimum value  ${\chi}_{\rm min}^2=1.66$.}
\label{fig:acc3}
\end{figure*}

\begin{figure*}
\centering
\includegraphics[height=9.7cm]{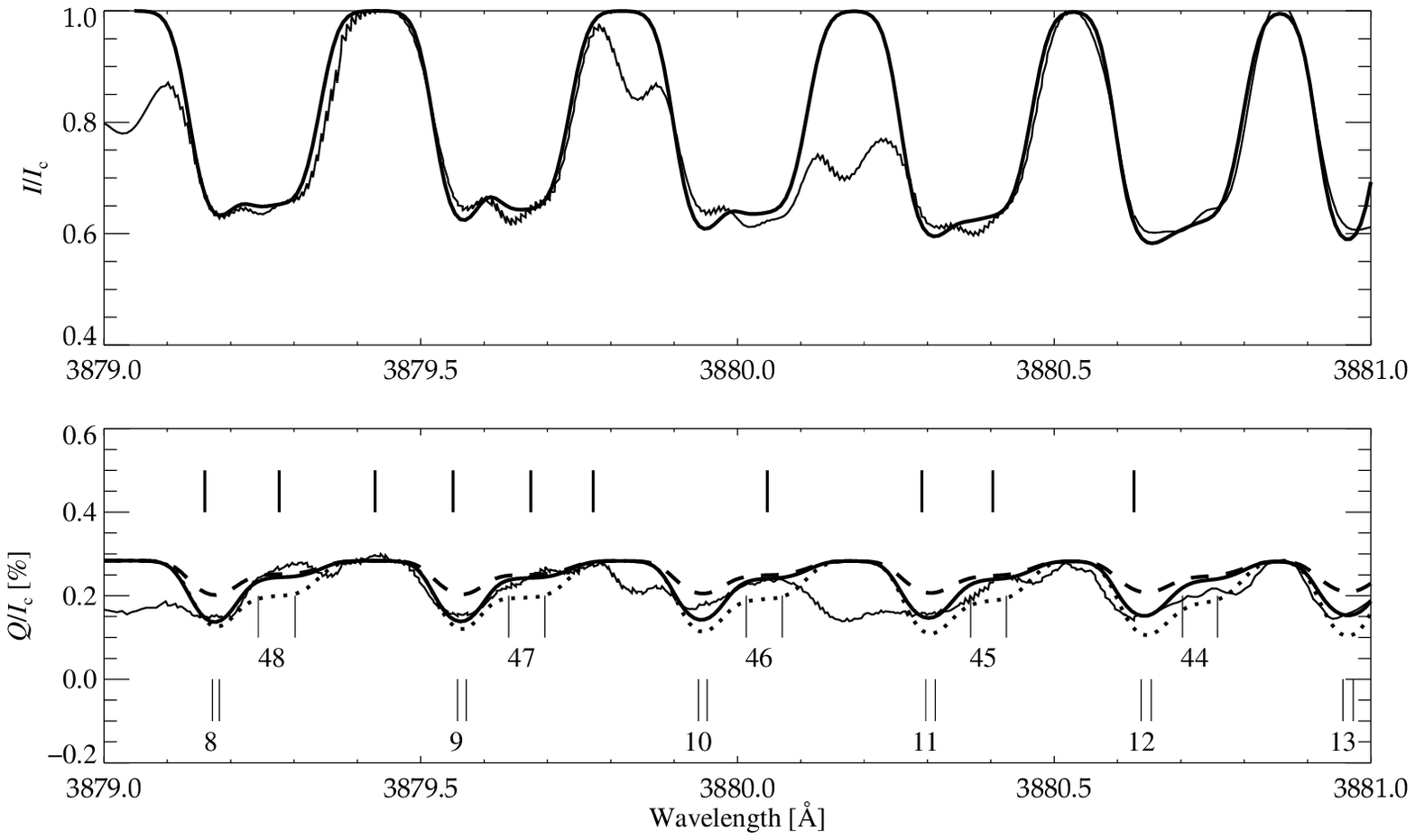}
\caption{Fits to the observations (thin solid  lines) of the CN violet system in the third wavelength region close to the (0,0) bandhead. The magnetic field strengths in the modeled spectra were set to 0 G (no depolarization, dashed line), 11 G (best fit, thick solid line), and 25 G (too much depolarization, dotted line). Labels at the bottom of the lower panel indicate the CN doublets with the $N$ quantum numbers of the lower states. The labels at the top of the lower panel indicate the wavelength points chosen for the fitting procedure.}
\label{fig:spectrum_ex3}
\end{figure*}

Figure \ref{fig:acc1} shows the accuracy of the fitted spectrum illustrated by $\chi^2$-contours within the parameter space. All four free parameters are well constrained by the data, and their best values are listed in Table~\ref{table:bestfit}. In particular we find a turbulent magnetic field strength of 27$\pm$5~G. The magnetic field $B$ and the scattering probability $c_\mathrm{cs}$ exhibit a somewhat linear dependence (Fig.~\ref{fig:acc1}, leftmost panel), which is easily understood. A greater scattering probability leads to additional line polarization that has to be compensated with a stronger Hanle effect and thus a larger magnetic field. Nonetheless, both $B$ and $c_\mathrm{cs}$ remain restricted to well defined limits because of the differential sensitivity to the Hanle effect of the involved CN lines and the roughly line independent scattering probability. Similarly, the parameters $\cal{T}_\mathrm{norm}$ and $T_\mathrm{eff}$ also show a tendency towards a linear dependence, because a greater effective temperature reduces the line depth in the intensity spectrum that can be balanced by additional line opacity.

The modeled spectra are compared with observations in Fig.~\ref{fig:spectrum_ex1}. For the calculated curves we have assumed the best values of the parameters $T_\mathrm{eff}$, $c_\mathrm{cs}$, and ${\cal T}_\mathrm{norm}$ as listed in Table~\ref{table:bestfit} while varying the magnetic field strength. The differential Hanle effect is clearly visible and constrains the magnetic field strength to well defined limits. It has to be stronger than 15~G to saturate the highly sensitive $N\!=\!7$ doublet. The upper limit for the field strength is defined by the only modest observed depolarizations of the lines with $N\!=\!52$ and 69. The observed doubling of the $N\!=\!52$ doublet probably appears due to a strong Ti I blend neglected in our model. We obtain the best fit for the magnetic field strength of 27~G.


\subsection{Region 2: diagnostics in the (1,1) bandhead}\label{subsec:region2}
Molecular bandheads have a high potential for successful differential Hanle effect diagnostics. They contain a mixture of lines with different $J$-numbers and correspondingly different magnetic field sensitivities. On the other hand, all these lines belong to the same rotational branch and vibrational band so that many of the relevant line formation parameters nearly coincide. Here we present the results of our fitting procedure in the (1,1) bandhead of the CN violet system. 

The fitting procedure is illustrated in Fig.~\ref{fig:acc2} by the $\chi^2$-contours. Due to numerous neglected blends from atoms and molecules other than CN the quality of the fit is worse than in the first wavelength region (Sect.~\ref{subsec:region1}). Furthermore, the noise level in this region is also a bit higher. Nonetheless, we determine a magnetic field strength of 25$\pm$5~G in good agreement with the first region.

The modeled spectra are compared to observations in Fig.~\ref{fig:spectrum_ex2}. The bandhead of the (1,1) band is clearly visible near 3871.5~\AA. The selected wavelength region also contains a few lines from the (0,0) band. Contrary to the first region, lines with different Hanle effect sensitivities can not be easily distinguished as all lines are blended with each other. Even when neglecting the magnetic field it is possible to fit the overall average degree of polarization. However, the exact shape of the $Q/I_\mathrm{c}$ spectrum can only be reproduced when accounting for a magnetic field strength within  well defined limits.

In this wavelength region we had to adopt a cosine of the heliocentric angle $\mu\!=\!0.12$, which still lies within the error limits of the observed data. In fact, we failed to obtain good fits in this region with $\mu\!=\!0.10$, because the continuum polarization in the incident radiation field was too large.


\subsection{Region 3: diagnostics close to the (0,0) bandhead}\label{subsec:region3}

The CN lines near the (0,0) bandhead are very strong and heavily blended. This results in a significant optical thickness larger than 1 at the (0,0) bandhead so that the single scattering approximation loses its validity. Therefore, we selected a region which is 4~\AA\ away from the bandhead.

This third wavelength region contains a mixture of lines with $N\!\sim\!10$ and $N\!\sim\!40$ from the P branch. The main features relevant for diagnosing the magnetic field in this spectral window are groups of neighboring lines containing a doublet with $N\!\sim\!10$ and one with $N\!\sim\!40$, which are particularly well seen in the intensity spectrum. The doublets within one group can however be distinguished due to a consistent wavelength shift and different polarization characteristics. The lines with $N\!\sim\!10$ have a larger effective polarizability $W_2\!\sim\!0.5$ (compared to $W_2$ close to the asymptotic limit of 0.1 for lines with $N\!\sim\!40$), which in principle causes a larger line polarization in a non-magnetic medium. The Hanle effect introduces a competing contribution to the relative line polarization because the lines with small $N$ are much more sensitive to the magnetic field. From Fig.~\ref{fig:spectrum_ex3} we see that the consistent modeling leads to smaller polarization and the expected stronger Hanle effect in the lines with small $N$. Therefore, we can indeed infer the magnetic field strength from the $Q/I_\mathrm{c}$ spectrum.

The results of the fitting procedure are presented in Figs.~\ref{fig:acc3} and \ref{fig:spectrum_ex3}. Again all the free parameters are well constrained and we reach a reasonable quality of the fit, except in three narrow regions due to missing blend lines in our model at 3879.0~\AA, 3879.85~\AA, and 3880.2~\AA, which is clearly seen in the intensity spectrum. The best fit is obtained for a magnetic field strength of 11$\pm$2~G, which is significantly smaller than in the first two analyzed spectral windows.


\subsection{Discussion}\label{subsec:comparison}

\begin{table}
\caption{Least squares fits to the  magnetic field strength $B$, the optical depth normalization coefficient ${\cal T}_\mathrm{norm}$, the scattering probability $c_\mathrm{sc}$, the temperature of the isothermal CN layer $T_\mathrm{eff}$ and cosine of the heliocentic angle $\mu$ in three considered regions of spectrum} 
\label{table:bestfit}
\centering 
\begin{tabular}{ c  c  c c  c  c} 
\hline
\hline
  Region &  B, [G]     & ${\cal T}_\mathrm{norm}$ &   $c_\mathrm{cs}$ & $T_\mathrm{eff}, K$  & $ \mu $ \\ 
\hline 
   1 &    $27 \pm 5$  &     2.0  &   0.375     &  4620  & 0.1   \\
   2 &    $25 \pm 5$  &     3.5  &   0.240     &  4620  & 0.12  \\
   3 &    $11 \pm 2$  &     2.3  &   0.142     &  4460  & 0.11  \\
\hline 
\end{tabular}
\end{table}

\onltab{2}{
\begin{table*}
\caption{Line parameters used for the fitting procedure in the first spectral region (3865.9--3866.8~\AA). The columns list the wavelength ($\lambda$), the line identification (ID), the polarizability ($W_2$), the lifetime of the upper level ($\tau_R$), the turbulent magnetic field strength at which the line polarization is reduced to 30\% of its non-magnetic value neglecting elastic collision ($B_{30}$), the vibrational number which coincides for the upper and lower states ($v$), and the oscillator strength ($f_{vJS}$). } 
\label{table:region1}
\centering 
\begin{tabular}{ c  c  c c  c  c c | c  c  c c  c  c c} 
\hline
\hline
 $\lambda$, \AA\  &  ~~~ID     & $W_2$ &   $\tau_R$,  $10^{-8}$\,s & $B_{30}$  & v  & $f_{vJS}$ &  $\lambda$, \AA\  &  ~~~ID     & $W_2$ &   $\tau_R$,  $10^{-8}$\,s & $B_{30}$  & v  & $f_{vJS}$ \\  
 \hline
3865.978&$   R_{1}(12)$&   0.1239&   7.2719&    22&   0&      0.391&3865.998&$   R_{2}(12)$&   0.1260&   7.2736&    22&   0&      0.362\\
3866.013&$   Q_{12}(12)$&   0.3982&   7.2736&    22&   0&      0.001&3866.096&$   P_{1}( 6)$&   0.0604&   7.4397&     9&   1&      0.154\\
3866.090&$   Q_{21}( 6)$&   0.3916&   7.4397&     9&   1&      0.002&3866.098&$   P_{2}( 6)$&   0.0545&   7.4635&     9&   1&      0.130\\
3866.160&$   P_{1}(69)$&   0.0957&   7.0063&   111&   0&      2.016&3866.086&$   Q_{21}(69)$&   0.3999&   7.0063&   111&   0&      0.000\\
3866.235&$   P_{2}(69)$&   0.0957&   7.0069&   111&   0&      1.987&3866.400&$   P_{1}(52)$&   0.0944&   7.2450&    86&   1&      1.254\\
3866.344&$   Q_{21}(52)$&   0.3999&   7.2450&    86&   1&      0.000&3866.460&$   P_{2}(52)$&   0.0943&   7.2456&    86&   1&      1.230\\
3866.540&$   P_{1}( 7)$&   0.0650&   7.4251&    11&   1&      0.178&3866.533&$   Q_{21}( 7)$&   0.3938&   7.4251&    11&   1&      0.002\\
\hline
\end{tabular}
\end{table*}
}

\onltab{3}{
\begin{table*}
\caption{Line parameters used for the fitting procedure in the second spectral region (in the (1,1) bandhead), with the same parameters as in Table~\ref{table:region1}.} 
\label{table:region2}
\centering 
\begin{tabular}{ c  c  c c  c  c c |  c  c  c c  c  c c} 
\hline
\hline
  $\lambda$, \AA\  &  ~~~ID     & $W_2$ &   $\tau_R$,  $10^{-8}$\,s & $B_{30}$  & v  & $f_{vJS}$ &  $\lambda$, \AA\  &  ~~~ID     & $W_2$ &   $\tau_R$,  $10^{-8}$\,s & $B_{30}$  & v  & $f_{vJS}$ \\  
\hline 
3869.672&$   P_{1}(16)$&   0.0829&   7.3832&    26&   1&      0.394&3869.655&$   Q_{21}(16)$&   0.3988&   7.3832&    26&   1&      0.001\\
3869.695&$   P_{2}(16)$&   0.0819&   7.3845&    26&   1&      0.370&3869.831&$   P_{1}(42)$&   0.0931&   7.2954&    70&   1&      1.015\\
3869.786&$   Q_{21}(42)$&   0.3998&   7.2954&    70&   1&      0.000&3869.886&$   P_{2}(42)$&   0.0929&   7.2959&    70&   1&      0.991\\
3869.920&$   R_{1}( 7)$&   0.1397&   7.2917&    14&   0&      0.246&3869.938&$   R_{2}( 7)$&   0.1457&   7.2980&    14&   0&      0.217\\
3869.947&$   Q_{12}( 7)$&   0.3953&   7.2980&    14&   0&      0.002&3869.916&$   P_{1}(17)$&   0.0838&   7.3806&    28&   1&      0.418\\
3869.898&$   Q_{21}(17)$&   0.3990&   7.3806&    28&   1&      0.001&3869.940&$   P_{2}(17)$&   0.0829&   7.3817&    28&   1&      0.394\\
3870.070&$   P_{1}(64)$&   0.0954&   7.0428&   103&   0&      1.871&3870.001&$   Q_{21}(64)$&   0.3999&   7.0428&   103&   0&      0.000\\
3870.146&$   P_{2}(64)$&   0.0953&   7.0433&   103&   0&      1.842&3870.070&$   P_{1}(41)$&   0.0929&   7.2999&    68&   1&      0.991\\
3870.026&$   Q_{21}(41)$&   0.3998&   7.2999&    68&   1&      0.000&3870.118&$   P_{2}(41)$&   0.0928&   7.3004&    68&   1&      0.967\\
3870.146&$   P_{1}(18)$&   0.0846&   7.3780&    29&   1&      0.442&3870.127&$   Q_{21}(18)$&   0.3991&   7.3780&    29&   1&      0.001\\
3870.176&$   P_{2}(18)$&   0.0838&   7.3790&    29&   1&      0.418&3870.282&$   P_{1}(40)$&   0.0928&   7.3043&    66&   1&      0.967\\
3870.239&$   Q_{21}(40)$&   0.3998&   7.3043&    66&   1&      0.000&3870.340&$   P_{2}(40)$&   0.0926&   7.3048&    66&   1&      0.944\\
3870.460&$   P_{1}(19)$&   0.0854&   7.3754&    31&   1&      0.466&3870.440&$   Q_{21}(19)$&   0.3992&   7.3754&    31&   1&      0.001\\
3870.489&$   P_{2}(19)$&   0.0846&   7.3763&    31&   1&      0.442&3870.530&$   P_{1}(39)$&   0.0926&   7.3086&    65&   1&      0.944\\
3870.488&$   Q_{21}(39)$&   0.3998&   7.3086&    65&   1&      0.000&3870.580&$   P_{2}(39)$&   0.0924&   7.3091&    65&   1&      0.920\\
3870.551&$   P_{1}(20)$&   0.0861&   7.3728&    33&   1&      0.489&3870.530&$   Q_{21}(20)$&   0.3992&   7.3728&    33&   1&      0.001\\
3870.582&$   P_{2}(20)$&   0.0854&   7.3736&    33&   1&      0.466&3870.657&$   R_{1}( 6)$&   0.1457&   7.2987&    12&   0&      0.217\\
3870.664&$   R_{2}( 6)$&   0.1538&   7.3078&    12&   0&      0.187&3870.672&$   Q_{12}( 6)$&   0.3938&   7.3078&    12&   0&      0.002\\
3870.715&$   P_{1}(21)$&   0.0867&   7.3701&    35&   1&      0.513&3870.693&$   Q_{21}(21)$&   0.3993&   7.3701&    35&   1&      0.001\\
3870.751&$   P_{2}(21)$&   0.0861&   7.3708&    35&   1&      0.489&3870.662&$   P_{1}(38)$&   0.0924&   7.3128&    63&   1&      0.920\\
3870.621&$   Q_{21}(38)$&   0.3998&   7.3128&    63&   1&      0.000&3870.715&$   P_{2}(38)$&   0.0922&   7.3133&    63&   1&      0.896\\
3870.797&$   P_{1}(63)$&   0.0953&   7.0498&   102&   0&      1.842&3870.729&$   Q_{21}(63)$&   0.3999&   7.0498&   102&   0&      0.000\\
3870.871&$   P_{2}(63)$&   0.0953&   7.0503&   102&   0&      1.813&3870.872&$   P_{1}(22)$&   0.0872&   7.3674&    36&   1&      0.537\\
3870.849&$   Q_{21}(22)$&   0.3994&   7.3674&    36&   1&      0.001&3870.906&$   P_{2}(22)$&   0.0867&   7.3681&    36&   1&      0.513\\
3870.817&$   P_{1}(37)$&   0.0922&   7.3170&    61&   1&      0.896&3870.777&$   Q_{21}(37)$&   0.3998&   7.3170&    61&   1&      0.000\\
3870.870&$   P_{2}(37)$&   0.0920&   7.3174&    61&   1&      0.872&3871.007&$   P_{1}(23)$&   0.0878&   7.3647&    38&   1&      0.561\\
3870.983&$   Q_{21}(23)$&   0.3994&   7.3647&    38&   1&      0.001&3871.042&$   P_{2}(23)$&   0.0872&   7.3653&    38&   1&      0.537\\
3870.956&$   P_{1}(36)$&   0.0920&   7.3210&    60&   1&      0.872&3870.917&$   Q_{21}(36)$&   0.3998&   7.3210&    60&   1&      0.000\\
3871.107&$   P_{2}(36)$&   0.0918&   7.3214&    60&   1&      0.848&3871.125&$   P_{1}(24)$&   0.0882&   7.3618&    40&   1&      0.585\\
3871.099&$   Q_{21}(24)$&   0.3995&   7.3618&    40&   1&      0.000&3871.159&$   P_{2}(24)$&   0.0878&   7.3624&    40&   1&      0.561\\
3871.222&$   P_{1}(25)$&   0.0887&   7.3589&    41&   1&      0.609&3871.195&$   Q_{21}(25)$&   0.3995&   7.3589&    41&   1&      0.000\\
3871.079&$   P_{1}(35)$&   0.0918&   7.3249&    58&   1&      0.848&3871.042&$   Q_{21}(35)$&   0.3998&   7.3249&    58&   1&      0.000\\
3871.258&$   P_{2}(25)$&   0.0882&   7.3594&    41&   1&      0.585&3871.298&$   P_{1}(26)$&   0.0891&   7.3559&    43&   1&      0.633\\
3871.270&$   Q_{21}(26)$&   0.3996&   7.3559&    43&   1&      0.000&3871.125&$   P_{2}(35)$&   0.0916&   7.3254&    58&   1&      0.824\\
3871.382&$   P_{2}(27)$&   0.0891&   7.3533&    45&   1&      0.633&3871.417&$   P_{2}(28)$&   0.0895&   7.3501&    46&   1&      0.657\\
3871.338&$   P_{1}(27)$&   0.0895&   7.3528&    45&   1&      0.657&3871.309&$   Q_{21}(27)$&   0.3996&   7.3528&    45&   1&      0.000\\
3871.182&$   P_{1}(34)$&   0.0916&   7.3287&    56&   1&      0.824&3871.145&$   Q_{21}(34)$&   0.3997&   7.3287&    56&   1&      0.000\\
3871.329&$   P_{2}(26)$&   0.0887&   7.3564&    43&   1&      0.609&3871.364&$   R_{1}( 5)$&   0.1538&   7.3083&    11&   0&      0.187\\
3871.372&$   R_{2}( 5)$&   0.1655&   7.3224&    11&   0&      0.158&3871.379&$   Q_{12}( 5)$&   0.3916&   7.3224&    11&   0&      0.002\\
3871.222&$   P_{2}(34)$&   0.0913&   7.3292&    56&   1&      0.800&3871.258&$   P_{1}(33)$&   0.0913&   7.3324&    55&   1&      0.800\\
3871.223&$   Q_{21}(33)$&   0.3997&   7.3324&    55&   1&      0.000&3871.298&$   P_{2}(33)$&   0.0910&   7.3329&    55&   1&      0.776\\
3871.393&$   P_{1}(30)$&   0.0905&   7.3430&    50&   1&      0.728&3871.361&$   Q_{21}(30)$&   0.3997&   7.3430&    50&   1&      0.000\\
3871.438&$   P_{2}(30)$&   0.0902&   7.3435&    50&   1&      0.705&3871.393&$   P_{1}(29)$&   0.0902&   7.3464&    48&   1&      0.705\\
3871.362&$   Q_{21}(29)$&   0.3996&   7.3464&    48&   1&      0.000&3871.438&$   P_{2}(29)$&   0.0898&   7.3469&    48&   1&      0.681\\
3871.382&$   P_{1}(28)$&   0.0898&   7.3496&    46&   1&      0.681&3871.352&$   Q_{21}(28)$&   0.3996&   7.3496&    46&   1&      0.000\\
3871.366&$   P_{1}(31)$&   0.0908&   7.3396&    51&   1&      0.752&3871.333&$   Q_{21}(31)$&   0.3997&   7.3396&    51&   1&      0.000\\
3871.417&$   P_{2}(31)$&   0.0905&   7.3401&    51&   1&      0.728&3871.329&$   P_{1}(32)$&   0.0910&   7.3361&    53&   1&      0.776\\
3871.295&$   Q_{21}(32)$&   0.3997&   7.3361&    53&   1&      0.000&3871.360&$   P_{2}(32)$&   0.0908&   7.3365&    53&   1&      0.752\\\hline
\end{tabular}
\end{table*}
}
\onltab{4}{
\begin{table*}
\caption{Line parameters used for the fitting procedure in the third spectral region (close to the (0,0) bandhead), with the same parameters as in Table~\ref{table:region1}.} 
\label{table:region3}
\centering 
\begin{tabular}{ c  c  c c  c  c c |  c  c  c c  c  c c} 
\hline
\hline
  $\lambda$, \AA\  &  ~~~ID     & $W_2$ &   $\tau_R$,  $10^{-8}$\,s & $B_{30}$  & v  & $f_{vJS}$ &  $\lambda$, \AA\  &  ~~~ID     & $W_2$ &   $\tau_R$,  $10^{-8}$\,s & $B_{30}$  & v  & $f_{vJS}$ \\  
\hline 
3879.174&$   P_{2}( 8)$&   0.0650&   7.3078&    12&   0&      0.217&3879.185&$   P_{1}( 8)$&   0.0686&   7.2987&    12&   0&      0.246\\
3879.177&$   Q_{21}( 8)$&   0.3953&   7.2987&    12&   0&      0.002&3879.247&$   P_{1}(48)$&   0.0939&   7.1431&    78&   0&      1.407\\
3879.195&$   Q_{21}(48)$&   0.3999&   7.1431&    78&   0&      0.000&3879.306&$   P_{2}(48)$&   0.0938&   7.1435&    78&   0&      1.378\\
3879.566&$   P_{1}( 9)$&   0.0716&   7.2917&    14&   0&      0.275&3879.557&$   Q_{21}( 9)$&   0.3963&   7.2917&    14&   0&      0.002\\
3879.580&$   P_{2}( 9)$&   0.0686&   7.2980&    14&   0&      0.246&3879.649&$   P_{1}(47)$&   0.0938&   7.1485&    77&   0&      1.378\\
3879.598&$   Q_{21}(47)$&   0.3999&   7.1485&    77&   0&      0.000&3879.707&$   P_{2}(47)$&   0.0937&   7.1489&    77&   0&      1.349\\
3879.952&$   P_{1}(10)$&   0.0740&   7.2863&    16&   0&      0.304&3879.942&$   Q_{21}(10)$&   0.3970&   7.2863&    16&   0&      0.001\\
3879.967&$   P_{2}(10)$&   0.0716&   7.2909&    16&   0&      0.275&3880.029&$   P_{1}(46)$&   0.0937&   7.1538&    75&   0&      1.349\\
3879.980&$   Q_{21}(46)$&   0.3999&   7.1538&    75&   0&      0.000&3880.087&$   P_{2}(46)$&   0.0935&   7.1542&    75&   0&      1.320\\
3880.317&$   P_{1}(11)$&   0.0761&   7.2819&    17&   0&      0.333&3880.305&$   Q_{21}(11)$&   0.3975&   7.2819&    17&   0&      0.001\\
3880.332&$   P_{2}(11)$&   0.0740&   7.2854&    17&   0&      0.304&3880.389&$   P_{1}(45)$&   0.0935&   7.1590&    73&   0&      1.320\\
3880.340&$   Q_{21}(45)$&   0.3999&   7.1590&    73&   0&      0.000&3880.446&$   P_{2}(45)$&   0.0934&   7.1594&    73&   0&      1.291\\
3880.662&$   P_{11}(12)$&   0.0778&   7.2782&    19&   0&      0.362&3880.649&$   Q_{21}(12)$&   0.3979&   7.2782&    19&   0&      0.001\\
3880.678&$   P_{2}(12)$&   0.0761&   7.2809&    19&   0&      0.333&3880.728&$   P_{1}(44)$&   0.0934&   7.1641&    72&   0&      1.291\\
3880.681&$   Q_{21}(44)$&   0.3998&   7.1641&    72&   0&      0.000&3880.784&$   P_{2}(44)$&   0.0933&   7.1645&    72&   0&      1.262\\
3880.985&$   P_{1}(13)$&   0.0794&   7.2749&    21&   0&      0.391&3880.971&$   Q_{21}(13)$&   0.3982&   7.2749&    21&   0&      0.001\\

\hline
\end{tabular}
\end{table*}
}

The above comparison of the modeled spectra to the observed data was performed independently in the three selected wavelength regions. The obtained sets of free parameters for each region are collected in Table~\ref{table:bestfit}. They agree quite well in spite of the simple modeling, but there exist some differences that should be addressed.

The magnetic field strength found in the third region lies significantly below the value determined for the first two spectral regions. This is however not necessarily a contradiction, because the analyzed data were not observed simultaneously and sampled in principle different spatial positions on the Sun. 

Furthermore, the line strength and accordingly the optical thickness changes  from the first region to the third one. This influences the formation heights of the CN lines, which leads to different physical conditions in the corresponding atmospheric layers. Lines in the third region are stronger than in the first two and therefore form higher in the atmosphere where the temperature is smaller. A comparison of the contribution functions calculated for the typical lines in each spectral window shows that the difference in temperatures is around 100 K, consistent with the difference of 160 K obtained with our least-squares method (see Table~\ref{table:bestfit}). Our results suggest that the magnetic field strength may decrease with height, but a detailed interpretation of these small differences lies outside the scope of the current model and will be clarified by future work.

In addition, the continuum normalization and polarization in the third region remains an issue due to broad wings of nearby strong atomic lines including the H8 line from the Balmer series. The intensity normalization has little effect on the inferred magnetic field strength because it affects both the theoretical and observed $Q/I_{\rm c}$ signal in the same way. However, the additional opacity from those nearby atomic lines would probably reduce the continuum polarization and slightly increase the magnetic field strength by a few Gauss. 

Overall, the physical parameters determined with the fit lie in a reasonable range. The small scattering probability is consistent with expectations and with a relatively small influence of non-LTE effects. The total optical thickness of the CN layer is close to but still smaller than 1 (maximal optical thicknesses in the considered regions are $0.26$, $0.5$ and $0.32$ for the first, second and third regions correspondingly) in accordance with our assumptions. Finally, the temperature $T_\mathrm{eff}$ of the CN layer around 4500--4600~K that is required for the best fits, corresponds well to the temperature in the upper solar photosphere where the CN lines are formed.

An interesting extension of our method is the interpretation of the center-to-limb variations of the continuum and line radiation field, as it was done for Sr I 4607 {\AA} line by \citet{stenfloetal1997}  and for the $ \rm C_2$ and $ \rm MgH$ molecular lines by \citet{faurobertarnaud2003} . Such observation for the CN violet system are not available yet, but could be very useful for constraining the atmospheric models.


\section{Conclusions}
We have employed the differential Hanle effect for the interpretation of observations of the CN violet system, which allowed us to determine the turbulent magnetic field strength in the upper solar photosphere. This expands the number of molecules used for Hanle effect diagnostics, which was so far only based on C$_2$ and MgH. The main conclusions can be summarized as follows:
\begin{itemize}
  \item We have developed a new model that can diagnose the turbulent magnetic field via the differential Hanle effect. It is applicable for optically thin lines, relies on simple and few assumptions, and consistently accounts for continuum polarization, depolarization of the continuum by line opacity, intrinsic line polarization by coherent scattering, and the Hanle effect.
	\item Lines from the CN violet system are well suited as a diagnostic tool based on the differential Hanle effect. They are very prominent in the second solar spectrum, and there exist several narrow spectral regions containing simultaneously lines with small and large $J$ quantum numbers. These lines exhibit different sensitivities to the Hanle effect, since the effective Land\'{e} factor is inversely proportional to $J$ in the CN violet system.
	\item We have identified three spectral windows that are ideal for Hanle effect diagnostics and contain a small number of atomic and molecular blends.
	\item In all three spectral regions we obtained a good quality of the fit and determined turbulent magnetic field strengths of 27$\pm$5~G, 25$\pm$5~G, and 11$\pm$2~G, respectively.
\end{itemize}

\begin{acknowledgements}
This work was supported by  SNF grants 20002--103696 and PE002--104552. SB acknowledges the EURYI
Award from the ESF.
\end{acknowledgements}


\bibliographystyle{aa}
\bibliography{8046}


\end{document}